\documentclass[AMA,Times1COL]{WileyNJDv5} 

\articletype{Article Type}%

\received{Date Month Year}
\revised{Date Month Year}
\accepted{Date Month Year}
\journal{Journal}
\volume{00}
\copyyear{2023}
\startpage{1}

\usepackage{mathrsfs}

\begin{document}

\title{Output Feedback Design for Parameter Varying Systems subject to Persistent Disturbances and Control Rate Constraints}

\author[geas]{Jackson G. Ernesto}
\author[geas,EAS]{Eug\^{e}nio B. Castelan}
\author[Concordia]{Walter Lucia}

\authormark{Ernesto \textsc{et al.}}
\titlemark{Output Feedback Design for Parameter Varying Systems subject to Persistent Disturbances and Control Rate Constraints}

\address[geas]{Graduate Program in Automation and Systems Engineering 
(P\'{o}sAutoma\c{c}\~{a}o/UFSC),
$88040$-$900$, Florian\'{o}polis, SC, Brazil.}  

\address[EAS]{Department of Automation and Systems (EAS/CTC/UFSC), (EAS), Universidade Federal de Santa Catarina (UFSC) , $88040$-$900$, Florian\'{o}polis, SC, Brazil.}

\address[Concordia]{Concordia Institute for Information Systems Engineering (CIISE), Concordia University, Montreal, QC, H3G-1M8, Canada}             

\corres{
\email{jackson.ernesto@posgrad.ufsc.br} 
\email{eugenio.castelan@ufsc.br}
\email{walter.lucia@concordia.ca}}



\abstract[Abstract]{
This paper presents a technique for designing output feedback controllers for constrained linear parameter-varying systems that are subject to persistent disturbances. Specifically, we develop an incremental parameter-varying output feedback control law to address control rate constraints, as well as state and control amplitude constraints.
The proposal is based on the concept of robust positively invariant sets and applies the extended Farkas' lemma to derive a set of algebraic conditions that define both the control gains and a robust positively invariant polyhedron that satisfies the control and state constraints. These algebraic conditions are formulated into a bilinear optimization problem aimed at determining the output feedback gains and the associated polyedral robust positively invariant region. The obtained controller ensures that any closed-loop trajectory originating from the polyhedron converges to another smaller inner polyhedral set around the origin in finite time, where the trajectory remains ultimately bounded regardless of the persistent disturbances and variations in system parameters. Furthermore, by including the sizes of the two polyhedral sets inside the objective function, the proposed optimization can also jointly enlarge the outer set while minimizing the inner one. Numerical examples are presented to demonstrate the effectiveness of our proposal in managing the specified constraints, disturbances, and parameter variations.
}


\maketitle

\renewcommand\thefootnote{}

\renewcommand\thefootnote{\fnsymbol{footnote}}
\setcounter{footnote}{1}

\section{Introduction}\label{sec:Introduction}

A fundamental control challenge stems from designing stabilizing output-feedback control laws for constrained disturbed systems \cite{tarbouriech2011stability, Blanchini15}. Such constraints often originate from physical or security bounds imposed on systems. Meanwhile, external disturbances may affect the system's states or output measurement. Fortunately, such disturbances are often naturally bounded in amplitude, and they can be handled in controlled systems.
Nevertheless, both constraints and disturbances must be considered during control law designs for stability and performance guarantees.
Furthermore, it is usual that only a subset of the state variables is available for feedback, enforcing output feedback control laws. 
Moreover, linear systems influenced by time-varying parameters, known as Linear Parameter-Varying (LPV) systems, represents a particularly interesting and important class of systems that can also be used to model some classes of nonlinear systems \cite{mohammadpour2012control}. For example, LPV models can represent nonlinear systems through quasi-LPV or Takagi-Sugeno (T-S) fuzzy models which provides a comprehensive framework for analyzing and synthesizing parameter-varying control laws for unconstrained \cite{Tanaka02} and constrained nonlinear control systems, see \textit{e.g.} \cite{KCLS:15}. 

Recent works used various techniques to design controllers for constrained LPV systems. 
For example, utilizing Model Predictive Control (MPC), or, when dealing with disturbances, its counterpart, Robust Model Predictive Control (RMPC) (see, for example, \cite{yang2016optimal, ping2021tube, ding2019output}). Another popular technique stems from Linear Matrices Inequalities (LMIs), that uses either Lyapunov stability conditions or positive invariance conditions to guarantee the system's stability (see, for instance, \cite{sereni2022stabilization,peixoto2021improved, oliveira2021exponential, da2003local, kose2002rate}).
Furthermore, in \cite{hempel2011output}, the concept of polyhedral invariant sets is used to ensure stability through output feedback control of an LPV system, taking into consideration state and control amplitude constraints.
In \cite{dorea2020robust}, the authors focus on state feedback control laws for Fuzzy T-S systems that are also subject to persistent disturbances; the observer-based output feedback control is considered in \cite{ITC:23}, but disregarding disturbances. The last two works leverage the concepts of (Robust) Positive Invariance and contractivity of polyhedral sets to form  bilinear optimization problems for the design processes. However, the proposed methods do not address the limitations on the variation of the control rate, which is a primary concern of the current study.
Additionally, regarding control rate constraints, \cite{bender2012output} and \cite{da2008dynamic} design dynamical output feedback controllers through LMI optimization problems for continuous and discrete-time systems, respectively, which are also subject to bounded disturbances and control amplitude limits. However, in these papers, only LTI systems are considered.
Moreover, in \cite{franklineEugenio2021} the authors used the concept of Robust Positive Invariant (RPI) polyhedral sets to propose  a novel bilinear optimization technique to design output feedback controllers for assymetrical state and control constrained discrete LTI systems subject to persistent disturbances.  Finally, in \cite{ECSC:21, ECSC:21alt}, positive invariance of polyhedral sets and bilinear programming also allowed to propose control  design techniques  for discrete-time LPV systems subject to state and control amplitude, and control rate constraints, but in the absence of disturbances.

Inspired by the previous works that utilize polyhedral set-invariance properties and bilinear optimization design techniques, this paper proposes a method for designing output feedback controllers for constrained LPV systems subject to persistent disturbances and control rate constraints.
To tackle the control rate constraints in the design process, we develop an incremental parameter-varying output feedback control law, which also allows for incorporating state and control amplitude constraints. A novel aspect of the proposed control law, when compared to previous works \cite{ECSC:21, ECSC:21alt}, is the introduction of an important degree of freedom through the feedback from the actual measured outputs. Additionally, all control gain matrices considered are assumed to be parameter-dependent, distinguishing this method from the simpler strategy used in the previous study \cite{ECL:24}, which employed a constant gain matrix associated with the actual output. Overall, the controller design aims to ensure that any closed-loop trajectory originating from a large RPI polyhedron that meets the constraints converges to a smaller inner polyhedral set around the origin in finite time, where the trajectory remains ultimately bounded regardless of the persistent disturbances and variations in system parameters. As in \cite{franklineEugenio2021}, the proposed solution employs the extended Farkas' lemma to derive sets of algebraic conditions that characterize the RPI property of a polyhedron along with an associated UB-set and  constraints satisfaction. To address the challenges posed by the obtained augmented parameter-varying closed-loop system, we introduce a novel notation for matrix polytopes and propose an alternative closed-loop system formulation. This new approach, combined with Farkas' lemma, facilitates the derivation of the necessary and sufficient algebraic conditions to characterize the RPI and UB properties of the considered polyhedral sets, as well as to algebraically determine the satisfaction of control rate constraints. These algebraic relations are framed into a bilinear optimization problem, which seeks to simultaneously determine the output feedback gains and the associated polyhedral RPI and UB sets. Notably, these sets do not necessarily have to be homothetic to each other, differently from previous works \cite{franklineEugenio2021} and \cite{ECL:24}. Consequently, a newly tailored weighted objective function is proposed to enlarge the outer RPI set while minimizing the inner UB set.

The remainder of this paper is organized as follows. The next section introduces a novel notation for describing matrix polytopes and recalls the extended Farkas' Lemma. Section 3 presents the constrained control problem and offers an alternative formulation of the augmented LPV closed-loop system using the matrix polytopes notation. 
In Section 4, we outline the main contributions of this work: {\em i)} the necessary and sufficient algebraic conditions that characterize the polyhedral set-invariance concept, along with the inclusions that ensure the  state, control, and control rate constraints; and {\em ii)} the proposed bilinear optimization design technique. 
Section 5 features two numerical examples that demonstrate the application of the design technique, including a simulation involving a two-tank system. The paper concludes in Section 6 with some final thoughts. Additionally, an appendix provides proofs of the proposed results.

NOTATION: 
The sets of real numbers, real-valued column vectors of dimension $n>0$ and real-valued matrices of dimension $m\times n,\,m,n>0$ are denoted with $\Re,$ $\Re^{n}$ and $\Re^{m\times n},$ respectively.
Given an invertible square matrix $M,$ $M^{-1}$ denotes its inverse.  $M$ is a non-negative matrix if all its entries, namely $M_{ij},$ are non-negative, i.e. $M_{ij} \geq 0,  \forall i,j$. The vectors ${\bf{0}}_p\in \Re^n,\, {\bf{1}}_p\in \Re^n$ denote columns vectors containing only zeros or ones in all the components. Given a vector $v\in \Re^{n_v},$ $v_k$ represents the value of $v$ at the discrete time instant $k \in Z_{+} \cong \{0,1,...\}.$ Any closed convex polyhedral set $\mathcal{P} \in \Re^n$, containing the origin in its interior, is represented by $\mathcal{P} = \{x\in \Re^n: Px \leq \phi\}$, with $P \in \Re^{l_p \times n}$ and $\phi \in \Re^{l_p}$ a positive vector.


\color{black}


\section{Preliminaries}\label{sec:Preliminaries}

\subsection{On matrix polytopes}

Let $ M(\beta) = \displaystyle{\sum_{i=1}^{n_v}} \beta_i M_i,$ with $M_i \in \Re^{m \times n}$ and $ \beta \in \mathcal S_\beta := \left\{\beta \in \Re^{n_v},\,\beta_i \geq 0 \,; \displaystyle{\sum_{i=1}^{n_v}} \beta_i = 1\, \right\}.$ Then, by defining $$ \mathcal M = \begin{bmatrix}
    M_1 &  \ldots & M_{n_v}
\end{bmatrix} \in \Re^{m \times n_vn} \,,\, \mathcal M^c = \begin{bmatrix}
    M_1 \\ \vdots \\ M_{nv}
\end{bmatrix} \in \Re^{n_v m \times n}  $$
one has \begin{equation} \label{eq:mbeta_alternative}
M(\beta) = \mathcal M \Gamma(\beta) = \Gamma'(\beta) \mathcal M^c,
\end{equation}
where $\Gamma'(\beta) = \begin{bmatrix}
    \beta_1 I, \ldots \beta_{n_v} I
\end{bmatrix} \in \Re^{ m\times n_vm}$ and $ \Gamma(\beta) \in \Re^{ n_vn \times n}$. Likewise, for $ N(\theta) = \displaystyle{\sum_{i=1}^{n_v}} \theta_j N_j,$ with $N_j \in \Re^{n \times p}$ and $ \theta \in \mathcal S_\theta := \left\{ \theta \in \Re^{n_v},\,\theta_j \geq 0 \,; \displaystyle{\sum_{j=1}^{n_v}} \theta_1 = 1\, \right\},$ it follows
\begin{equation} \label{eq:ntheta_alternative}
N(\theta) = \mathcal N \Gamma(\theta) = \Gamma'(\theta) \mathcal N^c,
\end{equation}
where $\mathcal N \in \Re^{ n \times n_vp }$, $\mathcal N^c \in \Re^{ n_vn\times p}$, $ \Gamma(\theta) \in \Re^{ n_v p\times p}$, and $ \Gamma^c(\theta) \in \Re^{ n \times n_vn}.$
Using, the previous alternative LPV notation, we have:

$\bullet$ \textit{Composed product of two matrix polytopes}:
\begin{equation}
    \label{eq:lpv-product}
    M(\beta)N(\theta) = \Gamma'(\beta) \mathcal M^c \mathcal N \Gamma(\theta),
\end{equation}
where, by definition,
$
\mathcal M^c \mathcal N =
\begin{bmatrix}
    M_1N_1 & \ldots & M_1N_{n_v} \\
    \vdots & \ddots & \vdots \\
    M_{n_v}N_1 & \ldots & M_{n_v}N_{n_v}
\end{bmatrix}
$.

$\bullet$ \textit{Pre- and Post-multiplication of a matrix polytope by a constant one}:
If $M \in \Re^{m \times n}$, then\begin{equation}
    \label{eq:cte-premultiply-LPV-1}
    MN(\theta) = \Gamma'(\theta) diag(M) \mathcal N^c,
\end{equation}
where $diag(M) \in \Re^{n_v m \times n_vn}$. Likewise, if $N \in \Re^{n \times p}$
\begin{equation}
    \label{eq:cte-postmultiply-LPV-1}
    M(\beta)N = \mathcal M diag(N) \Gamma(\beta),
\end{equation}
where $diag(N) \in \Re^{n_v n \times n_vp}$.

$\bullet$ \textit{Matrix polytope representation of constant matrices}:
For $M \in \Re^{m \times n}$, consider $\mathcal M = \begin{bmatrix}
    M & \ldots & M
\end{bmatrix} \in \Re^{n \times n_v m}$. Then \begin{equation}
    \label{eq:constant-matrix}
   M  =\Gamma'(\beta) \mathcal M  = \mathcal M^c  \Gamma(\theta),
\end{equation} with compatible $\Gamma'(\beta)$ and $ 
\Gamma(\theta)$. In particular, the $n-$ dimensional identity matrix, $I \in \Re^{n \times n}$ may read a
$$ I := I(\beta) = \Gamma'(\beta) \mathcal I^c \text{ or } I := I(\theta) = \mathcal I \Gamma(\theta),$$ 
where $\mathcal I = \mathcal I^{c'}  = \begin{bmatrix}
    I & \ldots & I
\end{bmatrix} \in \Re^{n \times n_v n}$.

$\bullet$ \textit{ Sum $(\oplus)$ of a single matrix polytope with the composed product of two matrix polytopes}:

Let $ F(\theta) = \displaystyle{\sum_{j=1}^{n_v}} \theta_j F_j,$ with $F_j \in \Re^{m \times p}$ and $ \theta \in \mathcal S_\theta$. Then
\begin{equation} \label{eq:composed_sum}
\begin{array}{rcl}
S(\beta,\theta) &= & F(\theta)\oplus M(\beta)N(\theta):= I(\beta)F(\theta)+M(\beta)N(\theta) \\
 & = & \Gamma'(\beta) (\mathcal{I}^c \mathcal{F} +\mathcal{M}^c\mathcal{N})\Gamma(\theta) = \Gamma'(\beta) \mathcal S  \Gamma(\theta), 
\end{array}
\end{equation}
where $\mathcal S \in \Re^{n_v n \times n_v p }$, with
$$\mathcal S = 
\begin{bmatrix}
   F_1 + M_1N_1 & \ldots &  F_{n_v} +M_1N_{n_v} \\
    \vdots & \ddots & \vdots \\
    F_1 + M_{n_v}N_1 & \ldots & F_{n_v}+ M_{n_v}N_{n_v}
\end{bmatrix} := \begin{bmatrix}
    \mathcal S_{ij}
\end{bmatrix}.
$$

$\bullet$ \textit{Pre- and Post-multiplication of a composed matrix polytope by a constant one}:

If $M \in \Re^{m \times n}$, then\begin{equation}
    \label{eq:cte-premultiply-LPV}
    M S(\beta,\theta) = \Gamma'(\beta) diag(M) \mathcal S \Gamma(\theta),
\end{equation}
where $diag(M) \in \Re^{n_v m \times n_vn}$. Likewise, if $N \in \Re^{n \times p}$
\begin{equation}
    \label{eq:cte-postmultiply-LPV}
    S(\beta,\theta)N = \Gamma'(\beta) \mathcal S diag(N) \Gamma(\theta),
\end{equation}
where $diag(N) \in \Re^{n_v m  \times n_v p}$.

\subsection{Extended Farkas' lemma (EFL)}



 \begin{lemma}\label{lemma:efl} \cite{H:89, hennet95}
 Consider two polyhedral sets  of $\Re^{n}$, defined by $\mathcal{P}_i=  \{x\in \Re^{n} : P_ix \leq \phi_i\}$, for $i=1,2$,
with $P_i \in \Re^{l_{p_i} \times n_u}$ and positive vectors $\phi_i \in \Re^{l_{p_i}}$. Then
 $\mathcal{P}_1 \subseteq \mathcal{P}_2$ 
 or, equivalently, $P_2x \leq \phi_2$, $\forall x $ $:$ $P_1x \leq \phi_1$, if and only if there exists a non-negative matrix $Q \in \Re^{l_{p_2} \times l_{p_1} }$ such that $QP_1 = P_2$ and
 $Q\phi_1 \leq  \phi_2.$
 \end{lemma}

\section{Problem presentation} \label{sec:Problem}


Consider a linear parameter-varying (LPV) discrete-time system given by

\begin{subequations}\label{eq:systemplant}
\begin{align}
x_+ & =  A(\alpha)x + B(\alpha)u + B_p(\alpha)p \label{cba24_eq:plant-model} \\
y & =  C x + D_\eta \eta 
\end{align}
\end{subequations}
where, for $k \in \mathbb N$, $x \cong  x_k \in \Re^{n_x}$ is the state vector and $x_+ \cong x_{k+1}$, 
$u \cong u_k\in\Re^{n_u}$ is the control input, $ y \cong y_k \in \Re^{n_y}$ is the measured output vector, and $ p \cong p_k \in \Re^{n_p} $ and $\eta \cong \eta_k\in \Re^{n_\eta}$ are exogenous and bounded process and measurement disturbance vectors, respectively. The matrices in system \eqref{eq:systemplant} are such that 
$C \in \Re^{ n_y \times n_x}$,
$D_\eta \in \Re^{ n_y \times n_\eta}$,
and
\begin{equation} \label{eq:polytope}
\begin{bmatrix} A(\alpha) & B(\alpha) & B_p (\alpha) \end{bmatrix}
= 
\sum_{i=1}^{n_v} \alpha_{i,k}\begin{bmatrix} A_i &  B_i & B_{pi} \end{bmatrix},
\end{equation}
 with $A_i \in \Re^{{n_x} \times {n_x}}$, $B_i \in \Re^{n_x \times n_u}$ and $B_{pi} \in \Re^{n_x \times n_p}$,  for $i = 1, \ldots, {n_v}$, where the parameter-varying vector $\alpha \cong \alpha_{k} \in \mathcal{S} = \{ \alpha \in \Re^{n_v} : \alpha_{i} \geq 0, \sum_{i=1}^{n_v} \alpha_{i} = 1 \}$ is supposed to be available in real-time. 

 Moreover, the system is subject to state, control amplitude, and control rate variation constraints represented by the closed polyhedral sets:
\begin{subequations}
\label{cba_eq:constraints_sets}
\begin{align}
  \mathcal{X} = & \{x : Xx \leq \textbf{1$_{l_x}$}\},  X \in \Re^{l_x \times n_x}, \label{eq:state_cst} \\
  \mathcal{U} = & \{u : U u \leq \textbf{1$_{l_u}$}\},  U \in \Re^{l_u \times n_u}, \label{eq:control_cst} \\
  \mathcal{U}_\delta = & \{\delta u : U_\delta \delta u \leq \textbf{1$_{l_d}$}\},  U_\delta \in \Re^{l_{u_\delta} \times n_u}, \label{eq:deltau_cst} 
\end{align}
\end{subequations} 
where, by definition, $\delta u = u_+ - u$, and the persistent bounded disturbances
  \begin{subequations} \label{eq:distubancesset}
  \begin{align}
\mathcal{P} = & \{p :Pp \leq {\textbf{1}}_{l_p}\}, P \in \Re^{l_p \times n_p}, \label{eq:perturbset}\\
 \mathcal{N} = & \{\eta :N\eta \leq {\textbf{1}}_{l_n}\},  N \in \Re^{l_n \times n_\eta}. \label{eq:noiseset}
\end{align}
 \end{subequations} 

The desired control objective  is as follows:
\textit{Compute an incremental output feedback control law, possibly dependent of the varying parameters, 
\begin{equation}
 \label{eq:incremental_f}
    u_+ =  u + \overbrace{f(u,y,y_+,\alpha, \alpha_+)}^{\delta u \cong}, 
\end{equation}
 and an admissible set of initial conditions for the corresponding closed-loop system, denoted $\Lambda$, such that for any closed-loop initial state belonging to $\Lambda$, any persistent disturbances sequences $p \in \mathcal P$ and $\eta \in \mathcal N$, and for any varying parameters $(\alpha,\alpha_+) \in \mathcal S \times \mathcal S$, the corresponding closed-loop state trajectory obeys the state constraints, $x \in \mathcal{X}$, fulfills the  control amplitude and the control rate variation constraints,   $u \in \mathcal{U}$ and $\delta u \in \mathcal{U}_\delta$,
 and is ultimately bounded in a small set $\Lambda_0 \subseteq \Lambda$ around the origin.}

To pursue the control
objective, which, in particular, considers rate-control limits and an incremental output-like feedback control law, we formulate the problem from the definition of augmented state and output vectors, respectively, given by 
\begin{equation} \label{eq:augmented_states}
\xi = \begin{bmatrix}
        x' & u'
    \end{bmatrix}' \in\Re^{n_{\xi}},~  n_{\xi} = n_x +n_u
\end{equation} 
and 
\begin{equation} \label{eq:augmented_outpt}
\upsilon = \begin{bmatrix}
        y' & u' & y_+'
    \end{bmatrix}' \in\Re^{n_{\upsilon}},~  n_{\upsilon} = 2 n_y +n_u.
\end{equation}

Thus, we can define the following augmented LPV system from \eqref{eq:systemplant} such that the control variation vector $\delta u$ and the augmented output vector $\upsilon$, appear as  \textit{virtual} control input and output signals, respectively:
    \begin{subequations}
        \label{eq:augmented_system}
        \begin{align}
        \xi_+ & = \mathbb{A}(\alpha) \xi + \mathbb{B}\delta u + \mathbb{B}_{p}(\alpha) p \\
        \upsilon & = \mathbb{C} \begin{bmatrix}
            \xi \\
        x_+
         \end{bmatrix} + 
        \mathbb{D}_\eta \eta 
        \end{align}
\end{subequations}
    where 
$
    \mathbb{A}(\alpha) 
 =  \begin{bmatrix}
        A(\alpha) & B(\alpha) \\
            0 & I
        \end{bmatrix} , ~~    \mathbb{B}_p(\alpha) = \begin{bmatrix}
           {B}_{p} (\alpha) \\ 0
        \end{bmatrix} , ~~\mathbb{B} = \begin{bmatrix}
            0 \\ I
         \end{bmatrix},~~ \mathbb{C} = \begin{bmatrix}
            C & 0 & 0\\
            0 & I & 0 \\
            0 & 0 & C
        \end{bmatrix}$ and 
        $\mathbb{D}_\eta = \begin{bmatrix}
            D_\eta\\ 0 \\ 0
        \end{bmatrix}$.\\

Next, we can consider the following parameter-varying control increment input vector, which is the \textit{virtual} output feedback control input for the augmented system \eqref{eq:augmented_system},
\begin{equation}
\label{eq:contoutput}
  \delta u = \begin{bmatrix}
    K (\alpha)  & \bar{K} (\alpha) & \hat K(\alpha_+)
  \end{bmatrix} \begin{bmatrix}
     y \\ u \\ y_+
  \end{bmatrix} 
   = \mathbb{K}( \alpha_+,\alpha)\upsilon , 
\end{equation} 
where, by definition,
\begin{eqnarray*}
    \mathbb{K} (\alpha_+,\alpha)  =   \begin{bmatrix}
\left(\displaystyle{\sum_{i=1}^{n_v} }\alpha_{i}   \begin{bmatrix}
       K_i & \bar{K}_i
  \end{bmatrix}  \right)  &   \left(\displaystyle{\sum_{j=1}^{n_v} }\alpha_{+,j}~\hat K_j \right)
    \end{bmatrix}   =   \displaystyle{\sum_{i=1}^{n_v} }\displaystyle{\sum_{j=1}^{n_v} }\alpha_{i} \,\alpha _{+,j}~\begin{bmatrix} K_i & \bar{K}_i    &   \hat K_j
    \end{bmatrix}, 
    \end{eqnarray*}
with $K_i \in \Re^{n_u \times n_y}$, $\bar K_i \in \Re^{n_u \times n_u}$, $\forall i = 1,\ldots, {n_v}$, and $\hat K_j \in \Re^{n_u \times n_y}$.

 \begin{remark}\label{rem:incremental}
            Notice, from \eqref{eq:contoutput}, that the actual parameter-varying incremental control input, \eqref{eq:incremental_f}, to be applied to the plant \eqref{eq:systemplant} at each new discrete-time instant $k$, reads
\begin{equation} \label{eq:actual-control}
    u_{k} =  \left(I + \bar{K} (\alpha_{k-1})\right)u_{k-1} + 
    K (\alpha_{k-1})y_{k-1} + \hat K (\alpha_k) y_{k}. 
\end{equation}
In particular, in the initial instant $k=0$, $y_0$ is directly transferred to $u_0$, if $\hat K(\alpha_0) \neq 0$.
        \end{remark}

From \eqref{eq:augmented_system} and \eqref{eq:contoutput}, the closed-loop system can be represented by
\begin{equation}
\label{eq:clsystem}
    \xi_+ = \mathbb{A}^{cl}(\alpha_+,\alpha) \xi + \mathbb{B}^{cl}_d(\alpha_+, \alpha) d_+, 
\end{equation}
 where $d_+ = \begin{bmatrix}  p' & \eta' & \eta'_+ \end{bmatrix}' \in \Re^{n_d}, n_d = 2n_p + n_\eta$, and $$\mathbb{A}^{cl}(\alpha_+,\alpha) = \begin{bmatrix} A(\alpha) & B(\alpha) \\ K(\alpha) C + \hat K(\alpha_+) C A(\alpha) & I + \bar{K}(\alpha) + \hat K(\alpha_+) C B(\alpha)  \end{bmatrix}, \ \mathbb{B}^{cl}_{d}(\alpha_+,\alpha) = \begin{bmatrix} B_{p}(\alpha) &0 & 0 \\ \hat K(\alpha_+) C B_{p}(\alpha) & K(\alpha) D_\eta & \hat K(\alpha_+) D_\eta \end{bmatrix}.$$

Notice that, by definition, both sequences $\eta$ and $\eta_+$ are  bounded within the same set $\mathcal{N}$, \eqref{eq:noiseset}.
 Thus, from \eqref{cba_eq:constraints_sets}-\eqref{eq:distubancesset}, the closed-loop  system \eqref{eq:clsystem} is subject to  the control rate constraints represented by $\mathcal U_\delta$, as well as to the augmented state constraints represented by
\begin{equation} \label{eq:aug_state_cst}
      \Xi = \{\xi : \mathbb{X} \xi \leq \textbf{1}_{l_{\xi}}\}, ~ \mathbb L = \begin{bmatrix}
        X & 0 \\
        0 & U
    \end{bmatrix} \in \Re^{l_\xi \times n_{\xi}},
\end{equation}
where $l_\xi = l_x + l_u$, and the augmented persistent disturbance bounds 
\vspace*{-0.3cm}\begin{equation} \label{eq:aug_disturbances}
      \Delta = \{ d_+ : \mathbb{D} d_+ \leq \textbf{1}_{l_{\Delta}}\}, ~ \mathbb{D} = 
    \begin{bmatrix}
        P & 0 & 0\\
        0 & N & 0 \\
        0 & 0 & N
    \end{bmatrix}
    \in \Re^{l_d \times n_{d}},
    \end{equation}
where $l_d = l_p+2 l_\eta$.

Now, to determine the set of admissible initial augmented states such that the corresponding trajectories will  respect the constraints irrespective the applied bounded persistent disturbances, we introduce the concept of contractive Robust Positively Invariant (RPI) set (also called $\Delta$-invariant set), with a UB-set, extending to the parameter-varying augmented system \eqref{eq:clsystem} the Definition 1 in \cite[p. 9746]{franklineEugenio2021}.

\begin{definition}
\label{def:RPI} A set 
$\Lambda \in \Re^{n_\xi}$ is a contractive Robust Positive Invariant (RPI-) set of the system \eqref{eq:clsystem}, with ultimately bounded (UB-)set 
$ \Lambda^0 \subseteq \Lambda$, if for any initial condition $\xi_0 = \begin{bmatrix}
    x'_0 & u'_0
\end{bmatrix}' \in 
\Lambda$ 
and any disturbance sequence $ d_+ = \begin{bmatrix}
    p' & \eta' & \eta'_+
\end{bmatrix}' \in \Delta$, the corresponding state trajectory remains inside 
$\Lambda$, converge to $\Lambda^0$
in a finite number of steps, and remains ultimately bounded  within 
for all $(\alpha,\alpha_+) \in \mathcal S \times \mathcal S$.
\end{definition}

Hence, the control objective tackled in this work can be formulated within the augmented state framework, utilizing the above concept of RPI-set, as follows.

\begin{problem}\label{problem:single_set}
Find stabilizing control increment gains $K(\alpha)$, $\bar K(\alpha)$, and $\hat K(\alpha_+)$ in \eqref{eq:contoutput}, a {\em large} contractive RPI set $\Lambda \subseteq \Xi \in \Re^{l_\xi}$, thus verifying  the augmented state constraints \eqref{eq:aug_state_cst}, with a {\em small} UB-set $\Lambda^0 \subseteq \Lambda$,  such that, for any initial condition $\xi_0 \in \Lambda$, $d_k \in \Delta$, and for all $(\alpha, \alpha_+ ) \in \mathcal S \times \mathcal S$,  
the control rate variation constraint \eqref{eq:deltau_cst} is also  fulfilled.
\end{problem}

\subsection{Alternative LPV formulation} \label{sec:alternative}

In the next section, we use the following lemma and its corollary, which proofs appear in the Appendix, to establish the results that base our solution to  Problem \ref{problem:single_set}. They reformulate the closed-loop system and control increment dynamics using the LPV notation introduced in the Preliminaries,  yielding  to  describe algebraically and prove  the desired closed-loop properties.

\color{black}
\begin{lemma} \label{lem:cl-loop-Big}
    The closed-loop system \eqref{eq:clsystem} can be equivalently re-written as
\begin{equation}
\label{eq:clsystemEquivalent}
\xi_+= \Gamma'(\alpha_+) \mathcal{A}^{cl}\Gamma(\alpha) \xi(k)  + \Gamma'(\alpha_+) \mathcal{B}^{cl} \Gamma(\alpha)  d_+,
\end{equation}
 where 
%
$$
\mathcal{A}^{cl}= \begin{bmatrix}
 \mathcal{A}^{cl}_{ij}
\end{bmatrix} \cong 
\left[
\begin{array}{ccc}
  \mathcal{A}^{cl}_{11}   &  \ldots & \mathcal{A}^{cl}_{1n_v} \\
  \vdots& \ddots& \vdots\\
   \mathcal{A}^{cl}_{n_v1}   &  \ldots & \mathcal{A}^{cl}_{n_vn_v}
\end{array}
\right] \in R^{(n_v n_{\xi}\times n_v n_{\xi})}, ~~ \mathcal{B}^{cl}= \begin{bmatrix}
 \mathcal{B}^{cl}_{ij}
\end{bmatrix} \cong
\left[
\begin{array}{ccc}
  \mathcal{B}^{cl}_{11}   &  \ldots & \mathcal{B}^{cl}_{1n_v} \\
  \vdots& \ddots& \vdots\\
   \mathcal{B}^{cl}_{n_v1}   &  \ldots & \mathcal{B}^{cl}_{n_vn_v}
\end{array}
\right] \in R^{(n_v n_{\xi}\times n_vn_d )},
$$
with, by definition,%
$$\mathcal{A}^{cl}_{i,j} = 
    \begin{bmatrix}
        A_i & B_i \\
        (K_i C + \hat K_j C A_i) & (I + \bar{K}_i + \hat K_j C B_i) 
    \end{bmatrix}   \text{ and } 
  \mathcal{B}^{cl}_{i,j} = 
    \begin{bmatrix}
        B^p_i &0 & 0 \\
        \hat K_j C B^p_i & K_i 
        D_\eta & \hat K_j D_\eta
    \end{bmatrix},\quad \forall\,(i,j) = 1, \ldots, n_\nu.
    $$
\end{lemma}
\textbf{Proof:} See the Appendix A. $~\Box$

Next, from \eqref{eq:clsystem}, the control increment \eqref{eq:contoutput} reads
\begin{equation}
 \label{eq:incrementFG}
    \delta u =  \mathbb A^{\delta_u}(\alpha,\alpha_+)\xi + \mathbb B^{\delta_u}(\alpha,\alpha_+) d_+,
\end{equation}
 where 
$\mathbb A^{\delta_u}(\alpha,\alpha_+) = \begin{bmatrix}
    K(\alpha) C + \hat K(\alpha_+) C A(\alpha) & \bar{K}(\alpha) + \hat K(\alpha_+) C B(\alpha)  \end{bmatrix}, ~~ 
    \mathbb B^{\delta_u}(\alpha,\alpha_+) = \begin{bmatrix}
    \hat K(\alpha_+) C B_{p}(\alpha) & K(\alpha) D^\eta & \hat K(\alpha_+) D^\eta \end{bmatrix}.$ \\
    
It leads to the following corollary of Lemma \ref{lem:cl-loop-Big}.

\begin{corollary}
\label{lem:deltau-Big}
The control increment \eqref{eq:contoutput} 
can be equivalently re-written as
\begin{equation}
\label{eq:deltauEquivalent} \delta u = \Gamma'(\alpha_+)\mathcal{A}^{\delta_u}\Gamma(\alpha)\xi +  \Gamma'(\alpha_+) \mathcal{B}^{\delta_u}\Gamma(\alpha) d_+, \end{equation}
where  
$\mathcal{A}^{\delta_u} = \begin{bmatrix}\mathcal{A}^{\delta_u}_{i,j}\end{bmatrix} \in \Re^{n_v n_u \times n_v n_\xi}$ and $\mathcal{B}^{\delta_u} = \begin{bmatrix} \mathcal{B}^{\delta_u}_{i,j} \end{bmatrix} \in \Re^{n_v n_u \times n_v n_d}$,   
%
with
$$\mathcal{A}^{\delta_u}_{i,j} = 
    \begin{bmatrix}
        (K_i C + \hat K_j C A_i) & (\bar{K}_i + \hat K_j C B_i) 
    \end{bmatrix}
    \text{ and } \mathcal{B}^{\delta_u}_{i,j} = 
    \begin{bmatrix}
        \hat K_j C B^p_i & K_i D^\eta & \hat K_j D^\eta
    \end{bmatrix},\quad \forall\,(i,j) = 1, \ldots, n_\nu.
    $$
\end{corollary} 
\textbf{Proof:} See the Appendix A. $~\Box$

\section{Main results} \label{sec:solution}
To tackle {\it Problem 1}, we first define the polyhedral sets:
%
\begin{subequations}
\label{setsL}
\begin{align}
\Lambda & = \{\xi: \mathbb L \xi \leq  \textbf{1}_l\},
\label{eq:setl}\\
\Lambda^0& =  \{\xi: \mathbb L \xi \leq  \boldsymbol{\rho} \}, \label{eq:set2}
\end{align}    
\end{subequations}
where $ \mathbb L \in \Re^{l_r \times n_\xi}, ~ l_r > n_\xi, rank(\mathbb L) = n_\xi, $ {and} the non-negative vector $\boldsymbol{\rho}  = \begin{bmatrix}
    \rho_1 \ldots \rho_{l_r}
\end{bmatrix}' \in \Re^{l_r} $ verifying  $ \mathbf{0} \leq \boldsymbol{\rho} \leq \mathbf{1}_{l_r}$, which guarantees $\Lambda^0 \subseteq \Lambda$. Note that $\Lambda^0$ is not necessarily a homotetic set of $\Lambda$, since each  face may be scaled by different values for $\rho_i \in [0\,,\,1]$.  As in \cite{franklineEugenio2021}, $l_r$ defines the {\em set complexity} for both  $\Lambda $ and $\Lambda^0$. Moreover, the matrix $\mathbb L = \begin{bmatrix} L_x & L_u \end{bmatrix}$ can be composed by $L_x \in \Re^{l_r \times n_x}$ and $L_u \in \Re^{l_r \times n_u}$.

\subsection{RPI algebraic conditions} \label{susec:RPI}

From Definition \ref{def:RPI}, we propose the following  necessary and sufficient algebraic characterization of the RPI property of the polyhedral set $\Lambda$, with UB-set $\Lambda^0$. The proof is divided into three parts, and it is shown in the Appendix. In the two first parts, the EFL and the alternative notation used to represent the closed-loop system by \eqref{eq:clsystemEquivalent}, Lemma \ref{lem:cl-loop-Big},  play a key role to obtain the proposed finite-dimensional invariance relations from  the infinite-dimensional characterization of the RPI property.

\begin{theorem}
    \label{prop:RPI-conditiona}
Consider the LPV system \eqref{eq:systemplant} and the incremental control \eqref{eq:incremental_f}, with $\delta_u$ given by \eqref{eq:contoutput}. Then,  the polyhedron $\Lambda$ in \eqref{eq:setl}, is a contractive RPI set of the closed-loop system \eqref{eq:clsystem}, with UB-set $\Lambda^0$ given by \eqref{eq:set2}, if and only if there exist non-negative matrices $\mathcal H \in \Re^{n_vl_r \times n_vl_r}$ and $\mathcal V \in  \Re^{n_vl_r \times n_vl_d}$,  a positive vector $\boldsymbol{\rho} \leq \mathbf{1}_{l_r} $ and a real scalar $\lambda \in [0 ,1)$,,
such that:
\begin{subequations}\label{const}
\begin{align}
\mathcal H\,diag(\mathbb L) & =  diag(\mathbb L)\,\mathcal A^{cl}, \label{nonlinear1}\\
\mathcal V\,diag(\mathbb D) &= diag(\mathbb L)\,\mathcal B^{cl}, \label{cond4} \\
\mathcal H\,diag(\mathbf{1}_{l_r}) + \mathcal V\,diag(\textbf{1}_{l_d}) &\leq \lambda\,\mathcal I^c \mathcal I\,diag(\textbf{1}_{l_r}), \label{cond1} \\
\mathcal H\,diag(\boldsymbol{\rho}) + \mathcal V\,diag(\mathbf 1_{l_d}) &\leq \epsilon_1\,\mathcal I^c\mathcal I\,diag(\boldsymbol{\rho}), \label{cond2}
\end{align}    
\end{subequations}
where the real positive scalar $ \epsilon_1 < 1$ is  sufficiently close to one.
\end{theorem}
\textbf{Proof:} See the Appendix A. $~\Box$

\subsection{Constraints fulfilment} \label{subsec:constraints}

Now, we resort the extended Farka's Lemma (EFL) 
for describing algebraically the inclusion of the RPI polyhedron $\Lambda$ in the set of extended state constraints $ \Xi$, \eqref{eq:aug_state_cst}, and its admissibility with regard the control increment constraints represented by $\mathcal U_\delta$, \eqref{eq:deltau_cst}.

First, by following the results in Lemma \ref{lemma:efl}, the inclusion $\Lambda \subseteq \Xi$ is verified, or, equivalently,  $\xi \in \Xi$, for all $\xi \in \Lambda$, if and only if there exists a non-negative matrix $\mathcal G \in \Re^{l_\xi \times l_r}$, 
such that:
\begin{subequations}\label{stateset}
    \begin{align}
\mathcal G \mathbb L &= \mathbb X, \label{nonlinear2} \\
\mathcal G \textbf{1}_{l_r} & \leq \textbf{1}_{l_\xi}. \label{condlem2} 
\end{align}    
\end{subequations} 

     Next, by resorting to  \eqref{eq:incrementFG}, the following admissibility condition must hold true to guarantee that any closed-loop trajectory starting from the RPI polyhedron $\Lambda$ fulfils the control increment constraint $\mathcal U_\delta$: 
\begin{equation} \label{eq:deltaufulfil_1}
\begin{array} {r}
U_\delta \delta u :=
 U_\delta \begin{bmatrix}
     \mathbb{A}^{\delta u}(\alpha, \alpha_+) & \mathbb{B}^{\delta u}(\alpha, \alpha_+)
 \end{bmatrix}   \left[\begin{array}{l}
      \xi    \\
      d_k      
      \end{array}
\right] \leq  \mathbf{1}_{l_{d}} , \\
\forall~ \xi \text{ and } \eta \text{ such that } \begin{bmatrix}
   \mathbb L & 0 \\ 0 & \mathbb D
\end{bmatrix}\begin{bmatrix}
    \xi \\ d_k
\end{bmatrix} \leq \begin{bmatrix}
    \mathbf 1_{l_r} \\ \mathbf 1_{l_d}
\end{bmatrix}.
\end{array}
\end{equation}

\begin{lemma}
\label{lem:deltau-admissible}
 The admissibility condition \eqref{eq:deltaufulfil_1} is equivalent to the existence of  non-negative matrices 
$ \mathcal Q \in \Re^{l_\delta n_v \times l_r n_v}$ and  $ \mathcal T \in \Re^{l_\delta n_v \times l_d n_v}$, such that:
\begin{subequations}\label{admissibleset}
    \begin{align}
\mathcal Q\,diag(\mathbb L) & =  diag(U_\delta)\mathcal A^{\delta_u}, \label{nonlinear31} \\
\mathcal T\, diag(\mathbb D) & =  diag(U_\delta)\mathcal B^{\delta_u},  \label{nonlinear3} \\
\mathcal Q\, diag(\textbf{1}_{l_r}) + \mathcal T diag(\textbf{1}_{l_d} )& \leq \mathcal I^c \mathcal I\,diag(\textbf{1}_{l_{u_\delta}}),  \label{condlem3}
\end{align}    
\end{subequations}
with 
$\mathcal A^{\delta_u}$ 
and 
$\mathcal B^{\delta_u}$ given by Corollary \ref{lem:deltau-Big}.
%
\end{lemma}
\textbf{Proof:} See the Appendix A. $~\Box$

\color{black}
\subsection{Proposed solution}

The following Proposition characterizes the considered solutions to Problem \ref{problem:single_set}. Before, we recall that the matrix $\mathbb L \in \Re^{ l_r \times n_\xi }$, with $l_r < n_\xi $, which shapes the polyhedrons $\Lambda$ and $\Lambda^0$, has to be full column-rank. This occurs if and only if $\mathbb{L}$ admits a left pseudo-inverse $\mathbb J \in   \Re^{ n_\xi  \times  l_r}$ such that
\begin{equation} \label{eq:pseudoinverse}
\mathbb J \mathbb L = 
I_{n_\xi}.
\end{equation}


\begin{proposition}
\label{prop:solution}
{Consider a system represented by the LPV-system \eqref{eq:systemplant}-\eqref{eq:polytope}, with associated constraints and disturbance bounds \eqref{cba_eq:constraints_sets}-\eqref{eq:distubancesset}, Then,  for a pre-assigned  {\em set complexity} $l_r > n_\xi$, Problem \ref{problem:single_set} admits a solution composed by
$\left( \mathbb K(\alpha_+, \alpha), \Lambda, \Lambda^0 \right)$,  if and only if there exist a real scalar $\lambda \in [ 0\,,\,1 ) $, a vector  $\boldsymbol{\rho} \in [ \mathbf{0}\,,\,\mathbf{1}_{l_r}] $, matrices $\mathbb{L}$
and $\mathbb J$,
and nonnegative matrices $\mathcal H$,
$\mathcal V$, $\mathcal Q$, $\mathcal T$ and $\mathcal G$
such that the set of  algebraic conditions \eqref{const}, \eqref{stateset},  \eqref{admissibleset} and \eqref{eq:pseudoinverse} hold true.}
    
\end{proposition}

\textbf{Proof:} It consists of combining the results stated in the present section.
$\Box$



\begin{remark}[Unconstrained control rate]
If the LPV system \eqref{eq:systemplant} is not subject to the control rate constraint \eqref{eq:deltau_cst}, Proposition \ref{prop:solution} can be easily adapted to characterize admissible solutions for an instance of Problem \ref{problem:single_set} that considers only the state and control constraints \eqref{eq:state_cst} and \eqref{eq:control_cst}. For that, it suffices to consider only the proposed relations \eqref{const}, \eqref{stateset} and \eqref{eq:pseudoinverse}. 
\end{remark}

\begin{remark}[Robust solution and LTI systems] To  characterize robust solutions for LPV or uncertain systems, it is possible to adapt Proposition \ref{prop:solution} by considering the time-invariant control increment  $\delta u = \mathbb K \upsilon_k $, with $\mathbb K= \begin{bmatrix}
    K & \bar{K} & \hat{K}
\end{bmatrix} $. Such control increment also applies to LTI systems and enhances the solution suggested in  Remark 3 of \cite{franklineEugenio2021}, which corresponds to $\mathbb K= \begin{bmatrix}
    K & \bar{K} & 0
\end{bmatrix} $.
\end{remark}
\subsection{Bilinear optimization-based design}

To apply the results summarized by Proposition  \ref{prop:solution}, it is worth to emphasize that the algebraic relations in  \eqref{const}, \eqref{stateset}, \eqref{admissibleset} and \eqref{eq:pseudoinverse} present bi-linear terms, meaning there are products between the elements of the set of decision variables  $\Gamma=$ $\{$ $\mathbb L$, $\boldsymbol{\rho}$, $ K_i$, $\bar K_i$, $\hat K_j$,
$\mathcal{H}$, $\mathcal{V}$, $\mathcal{G}$, $\mathcal{Q}$, $\mathcal{T}$,  $\mathbb J \}$, where the control gains $K_i$, $\bar K_i$ and $\hat{K}_j$ are the matrix variables embedded into $\mathcal{A}^{cl}$, $\mathcal{B}^{cl}$, $\mathcal{A}^{\delta_u}$ and $\mathcal{B}^{\delta_u}$. 
In fact, we  notice such bilinear products in \eqref{const} involving $\{\mathcal{H}, \mathbb{L}\}$, $\{\mathbb L,K_i, \bar K_i, \hat K_j\}$ and $\{ \mathcal{H}, \boldsymbol{\rho}\}$, 
in \eqref{stateset} between $\{ \mathcal{G}, \mathbb L \}$, in \eqref{admissibleset} between $\{ \mathcal{Q}, \mathbb L \}$, and in \eqref{eq:pseudoinverse} between $\{ \mathbb J, \mathbb L \}$.

Moreover, the complementary objectives in seeking solutions to Problem \ref{prop:solution} are to enlarge the outer RPI set, $\Lambda$, and to shrink the inner UB set, $\Lambda^0$, as much as possible. In particular,  to  enlarge  the size of $\Lambda$ in given directions, we introduce the  following auxiliary inequalities
\begin{equation} \label{sol1:eq:gammaz_in_L}
    \gamma_{t} \mathbb{L} \psi_t \leq \textbf{1}_{l_r}, \gamma_t > 0,  t= 1, \ldots, \bar{t}.
\end{equation}
where $\gamma_{t}$
are real positive scaling factors associated to a given set $\Psi$ of $\bar{t}>0$ directions $\psi_{t}\in \mathbb{R}^{n_\xi},$ where
 \begin{equation}\label{sol1:eq:setZ}
    \Psi = \{ \gamma_{t} \psi_t, t= 1, \ldots, \bar{t} \}.
\end{equation} 
%
%
with $\psi_t = \begin{bmatrix} \psi_{x,t}^T & \psi_{u,t}^T \end{bmatrix}^T$,  $\psi_{x,t} \in \mathbb{R}^{n_x}$ and $\psi_{u,t} \in \mathbb{R}^{n_u}$. 

Thus, from Proposition \ref{prop:solution} and discussion in 
the previous paragraphs, we propose  the following bi-linear optimization problem to find solutions to the Problem  \ref{problem:single_set}: 

\begin{equation}\label{eq:optproblem}
\begin{aligned}
& \underset{\Gamma, \gamma_{t}}{\text{maximize}}
& &  \mathcal{J} = (1-\theta)\sum_{t=1}^{\bar{t}} \frac{\gamma_{t}}{\bar{t}} - \theta \sum_{\ell=1}^{l_r} \frac{\rho_\ell}{l_r}   \\
&  \text{subject to} 
& &  \eqref{const},\eqref{stateset}, \eqref{admissibleset}, \eqref{eq:pseudoinverse},  
\eqref{sol1:eq:gammaz_in_L},\\ 
&&&  \text{ and } \underline{\Gamma} \leq \Gamma \leq \overline{\Gamma},
\end{aligned}
\end{equation}
where
\begin{itemize}
 \item the parameters $l_r$, $t$, $\theta$ and $\psi_t$ are the designer choices;
    \item  the objective function weights, by choosing $\theta$, the maximization of the robust positive invariant set in the given directions $\psi_t$, through its associated average scaling factors $\gamma_t$, and the minimization of the ultimately bounded set, through the minimization of the average scaling variables $\rho_\ell$; and
    \item the additional constraints  on $\Gamma$ impose lower and upper bounds, represented by ($\underline{\Gamma}$, $\overline{\Gamma}$), that limit the search space of solutions. See \cite[Remark 4]{franklineEugenio2021}.
\end{itemize}

   Optionally, in \eqref{eq:optproblem}, $\psi_{u,t}$ can also be considered as a supplementary decision variable. In such case,  $\psi_{u,t}$ appears as additional degrees of freedom to allow for the enlargement of the projection of the positive invariant sets $\Lambda$ and $\Lambda_0$ in the system state-subspace $\Re^{n_x}$.

\section{Numerical examples} \label{sec:examples}

To show the effectiveness of the proposed output feedback controller design strategy, this section shows numerical results obtained considering three different constrained systems. The examples were solved using the KNITRO solver \cite{knitro06}, with the Interior/CG (barrier) algorithm, multi-start option, and the other solver's default settings, with the following additional elementwise constraints: 
$\mathcal{H}, \mathcal{V}, \mathcal{G}, \mathcal{Q}, \mathcal{T}, \gamma_t:  [0, 10^2]~$ and
$~K_i, \bar{K}_{i}, \hat{K}_{j}, \mathbb{L}: [-10^2, 10^2],
 \mathcal{J}: [-10^3, 10^3].$ It should be noted that KNITRO does not guarantee to find globally optimal solutions, however local minima are found upon convergence \cite{franklineEugenio2021}.
\subsection{Example 1}

For the first example, we consider the same LTI system as in \cite{franklineEugenio2021}. To this end, consider system \eqref{eq:systemplant} and constraints \eqref{cba_eq:constraints_sets}-\eqref{eq:distubancesset} represented by the following data, 
\begin{equation*}
A  =  \left[ \begin{array}{*2r}
  1 &   1   \\
   0 &  1  
\end{array}\right],\,
B = \left[ \begin{array}{c}
b\\
1  
\end{array} \right], B_p  = \left[ \begin{array}{c}
1 \\
1  
\end{array} \right],
C = \left[ \begin{array}{cc} 1 &  0 \end{array} \right], D_\eta = 1,
\end{equation*}
with state and control constraints $-1 \leq x_1 \leq 1.25$, $|x_2| \leq 1$ and $-0.8 \leq u \leq 1$, or equivalently $X= \left[ \begin{array}{crrr}
0.8 ~ & 0 & -1 & 0\\
0 & 1 & 0 & -1 
\end{array} \right]'$ and $U = \left[ \begin{array}{cc}
1 &  -1.25 \end{array} \right]'$. The persistent disturbances are $| p | \leq 0.1$ and $| \eta  | \leq 0.1$, or equivalently  $P= N= \left[ \begin{array}{cc}
 10 & -10 
\end{array} \right]'$.
\subsubsection{LTI system:}
 We choose $b = 2$, the set complexity $l_r = 9$, meaning that the polyhedral set can assume at most $9$ faces, as in \cite[Section 5.1.3]{franklineEugenio2021}. 
Differently from  \cite{franklineEugenio2021},   the pursued objective is enlarge  the
projection of the RPI set $\Lambda$ instead of the cut of this set in the system state-subspace. Thus, in the present design we consider $\bar{t}=8$, the chosen $ \psi_{x,t}$ directions pointing to the state constraint vertices and gradient  directions, and, to exploit new degrees of freedom, $-1 \leq \psi_{u,t} \leq 1$ as part of the decision variables. 
 
\begin{figure}
    \centering
    \includegraphics[width=0.5\linewidth]{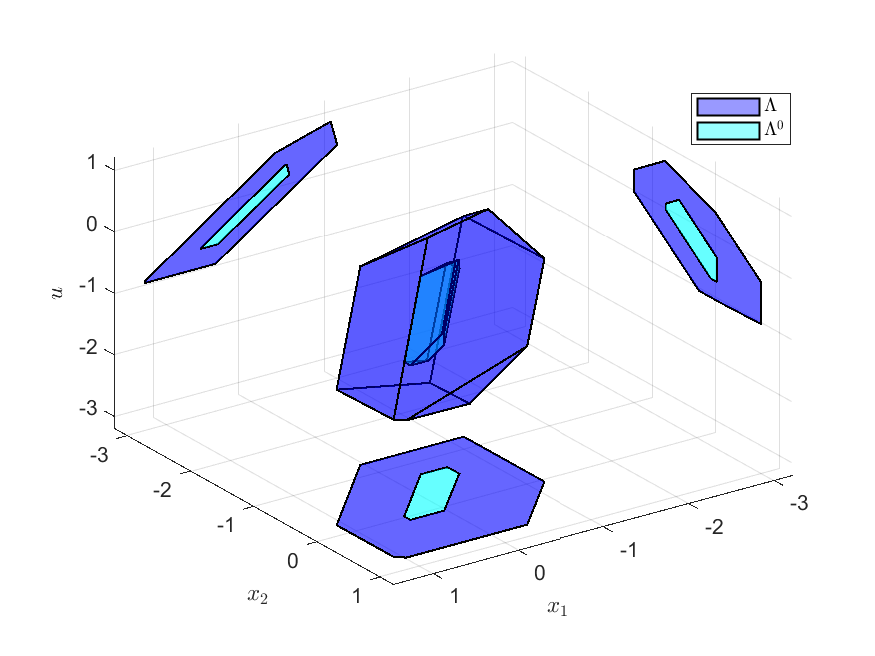}
    \caption{Example 1 - LTI, $\alpha = 0.5$}
    \label{fig:ex1_lti_alpha05}
\end{figure}

Table \ref{tab:LTI_dint} shows some design results obtained for different weights in the objective function, without considering control rate constraints.  As expected, the area of the $\Lambda$ projection  increases with smaller values of the weight $\theta$. Meanwhile, the $\Lambda^0$ Volume decreases with bigger valus of $\theta$. In Figure \ref{fig:ex1_lti_alpha05}, we showcase the results for $\theta = 0.5$, whith the RPI outter morst set, and the inner  UB-set. 
For comparative purposes, the biggest projection obtained in \cite[Section 5.1.3]{franklineEugenio2021} is $3.402$ vs $4.500$ using the technique proposed in this paper, demonstrating a $32\%$ increase in the projection area.


\begin{table*}[]
    \centering
    \begin{tabular}{|c||c|c|c|c|c|}
    \hline
        $\theta$ & $\Lambda$ Volume &$\Lambda$ Projection Area &  $\Lambda^0$ Volume & $\Lambda^0$ Projection Area & $[K ~~ \bar K ~~ \hat{K} ]$  \\ \hline
        0 & 1.1063 & 4.5000 & 1.0955 & 4.4684 & $\begin{bmatrix}
           0.4999 & -0.5000 & -0.6675 
        \end{bmatrix}$ \\
        0.1 & 1.8307 & 4.4999 & 0.3271 & 1.4166 & $\begin{bmatrix} 0.4999 & -0.5000 & -0.7765 \end{bmatrix}$ \\
        0.3 & 1.8307 & 4.4999 & 0.3271 & 1.4166 & $\begin{bmatrix} 0.4999 & -0.5000 & -0.7765 \end{bmatrix}$ \\
        0.5 & 1.3082 & 3.6966 &  0.0532 & 0.3832 & $\begin{bmatrix} 0.0000 & -1.0103 & -0.9747 \end{bmatrix}$ \\
        0.7 & 1.3349 & 3.2653 & 0.0429 & 0.3397 & $\begin{bmatrix}0.0000 & -0.9999 & -0.7226 \end{bmatrix}$ \\
        0.9 & 0.8282 & 2.2886 & 0.0323 & 0.2421 & $\begin{bmatrix}0.0000 & -1.0000& -0.7500 \end{bmatrix}$ \\
        1 & 0.8282 & 2.2886 & 0.0323 & 0.2421 & $\begin{bmatrix} 0.0000& -1.0000& -0.7500 \end{bmatrix}$ \\ \hline
    \end{tabular}
    \caption{Example 1  - Designs using different weights, $\theta$, for the constrained LTI system without control rate constraints}
    \label{tab:LTI_dint}
\end{table*}

\subsubsection{LPV system:}

In this second example we still consider the previous double integrator model with a varying parameter in the input matrix. It allows us to exploit some features of the design technique regarding the LPV model and the presence of control rate constraints. Thus, we let the input matrix to be parameter varying, $B(\alpha) = \begin{bmatrix} b & 1 \end{bmatrix}'$, by admitting $2 \leq b \leq 2.25$, and add the control rate  constraint  $-0.9 \leq \delta u \leq 0.6$, or equivalently $U_d = [1.6667 ~~ -1.1111]'$.  As in the first example, we consider the same $\psi_{t}$ directions and the set complexity $l_r = 9$. The results are shown in Table \ref{tab:dint_lpv} for different values of $\theta$. Once more, with lower values of $\theta$ we obtain bigger $\Lambda$ Projection Area, and with higher values of $\alpha$ we obtained smaller $\Lambda^0$ Volume.
\begin{figure}
    \centering
    \includegraphics[width=0.5\linewidth]{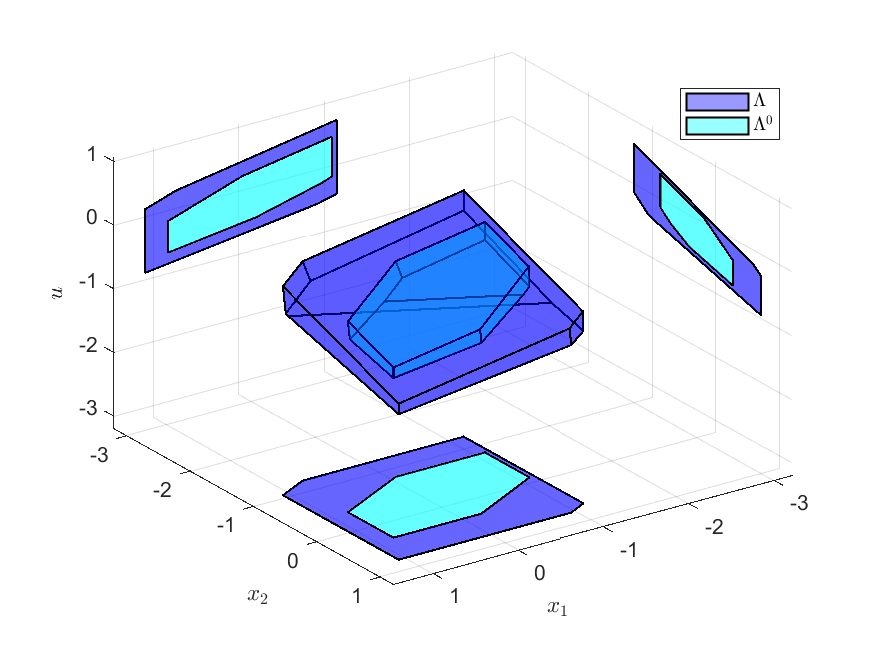}
    \caption{Example 1 - LPV, $\alpha = 0.5$}
    \label{fig:ex1_lpv_alpha05}
\end{figure}
Figure \ref{fig:ex1_lpv_alpha05} depicts the results for $\theta = 0.5$, where we present the $\Lambda$ RPI set and the $\Lambda_0$ UB-set.

Moreover, to showcase the effects of different choices of directions, we compare three different sets of directions, with $\theta = 0.5$, in Table \ref{tab:lpv_dir}. First, we show the results of choosing $\bar{t} = 4$ directions pointing towards the state constraint vertices, which gives the best result in terms of projection area. Next, $\bar{t}= 4$ directions are chosen as the gradient vectors of the state-constraint polyhedron. Lastly, we combined both choices made before, with a total of $\bar{t} = 8$ directions, where the result shows a good compromise between the projection area of $\Lambda$ and the volume of the UB-set $\Lambda_0$.

\begin{table*}[]
    \centering
    \begin{tabular}{|c||c|c|c|c|}
    \hline
        $\theta$ & $\Lambda$ Volume & $\Lambda^0$ Volume & $\Lambda$ Projection Area & $[K ~~\bar K ~~ \hat K]$\\ \hline
        0 & 1.44531 & 1.39625 & 4.49999 &
        $\begin{bmatrix}
         0.39495 & -0.52063 & -0.56621 \\
         0.38206 & -0.43701 & -0.56196
        \end{bmatrix}$  \\
        0.5 & 1.41657 & 0.48628 & 4.45545 &
        $\begin{bmatrix}
         0.39846 & -0.53465 & -0.57008 \\
         0.39741 & -0.43615 & -0.56948
        \end{bmatrix}$ \\
        1 & 0.52573 & 0.08852 & 1.97315 & 
        $\begin{bmatrix}
            0.00000 & -1.07292 & -0.63876 \\ 
            0.00000 & -0.98123 & -0.63876
        \end{bmatrix}$ \\ \hline
    \end{tabular}
    \caption{Example 1 - Designs using  different weights, $\theta$, for the constrained LPV system}
    \label{tab:dint_lpv}
\end{table*}

\begin{table*}[]
    \centering
\begin{tabular}{|c||c|c|c|c|}
    \hline
        Directions & $\Lambda$ Volume & $\Lambda^0$ Volume & $\Lambda$ Projection Area & $[K ~~\bar K ~~ \hat K]$\\ \hline
        Vertices of $\mathcal{X}$ & 1.43113 & 0.57721 & 4.49999 &
        $\begin{bmatrix}
        0.39819 & -0.53447 & -0.56970 \\
        0.39654 & -0.43622 & -0.56538
        \end{bmatrix} $ \\ \hline
        Gradients of $\mathcal{X}$ & 0.79426 & 0.50686 & 2.82688 & 
        $\begin{bmatrix}
        0.44588 & -0.55441 & -0.64347 \\
        0.44547 & -0.44244 & -0.64347 
        \end{bmatrix}$ \\ \hline
        Both & 1.41657 & 0.48628 & 4.45545 & 
        $\begin{bmatrix}
        0.39846 & -0.53465 & -0.57008 \\
        0.39741 & -0.43615 & -0.56948 
        \end{bmatrix}$ \\ \hline
    \end{tabular}
    \caption{Example 1 -  Designs using different  direction set choices, with fixed weight $\theta = 0.5$,  for the constrained LPV system}
    \label{tab:lpv_dir}
\end{table*}

\subsection{Example 2 - Coupled tank}

Next, we consider a two-tank system, specifically the digital twin (high fidelity simulator) provided by Quanser\textregistered. The system was modeled as a Quasi-LPV system subject to bounded disturbances, represents the nonlinear plant model inside a given polyhedral set $\mathcal X$. 
To showcase the potential of the proposed technique, we considered the shifted coupled tank system with an equilibrium point defined by $\hat{x}_{eq} = [15.25 ~14.87]'$ and $\hat{u}_{eq} = 8.1$, obtained experimentally, with the shifted states defined as $x = \hat{x} - \hat{x}_{eq}$ and the shifted control variable $u = \hat{u} - \hat{u}_{eq}$, resulting in the system in the form of \eqref{eq:systemplant} with vertex matrices
\begin{eqnarray*}
A_1 = \begin{bmatrix}
 0.9886 &0.0000 \\
 0.0112 &0.9886
\end{bmatrix},
 A_2 = \begin{bmatrix}
 0.9886 &0.0000 \\
 0.0112 &0.9840
\end{bmatrix}  \\
A_3 = \begin{bmatrix}
 0.9840 &0.0000 \\ 
 0.0158 &0.9886
\end{bmatrix}, 
A_4 = \begin{bmatrix}
 0.9840 &0.0000 \\
 0.0158 &0.9840
\end{bmatrix} 
\end{eqnarray*}

Additionally, the control input and to encompass control input uncertainties, originated from the pump gain variation, we considered $B = B_{p} = \begin{bmatrix}
 0.0179\\
 0.0001
\end{bmatrix}$, with the bounded disturbance $-0.256 \leq p \leq 0.256$.
Moreover, we considered the output matrices as $C = D_\eta = \begin{bmatrix} 1 &0 \\ 0 &1 \end{bmatrix} = I$. The systems constraints are $-5 \leq x_i \leq 5$ for $i = 1,2$, $-4 \leq u \leq 4$, $-2 \leq \delta u \leq 2$, and $-0.02 \leq \eta_i \leq 0.02$ for $i=1,2$, from which it is possible to define the constraint matrices $X = \begin{bmatrix} -0.2 & 0.2 & 0 & 0 \\ 0 & 0 & -0.2 & 0.2 \end{bmatrix}'$, $U= \begin{bmatrix} -0.25 & 0.25 \end{bmatrix}'$, $U_d = \begin{bmatrix} -0.5 & 0.5 \end{bmatrix}'$, and the disturbances matrices  $P = \begin{bmatrix} -3.9 & 3.9 \end{bmatrix}'$ and $N = \begin{bmatrix}
    -50 & 50 & 0 & 0 \\ 0 & 0 & -50 & 50
\end{bmatrix}'$. Notice, in particular, that the state constraints correspond to the bounds considered in the fuzzification process of the original system, meaning that the corresponding sets $\mathcal{X}$ and $\mathcal{U}$ of state and control amplitude constraints defines the validity for the considered Fuzzy T-S model \cite{KCLS:15}. 

\begin{figure}
    \centering
    \includegraphics[width=0.5\linewidth]{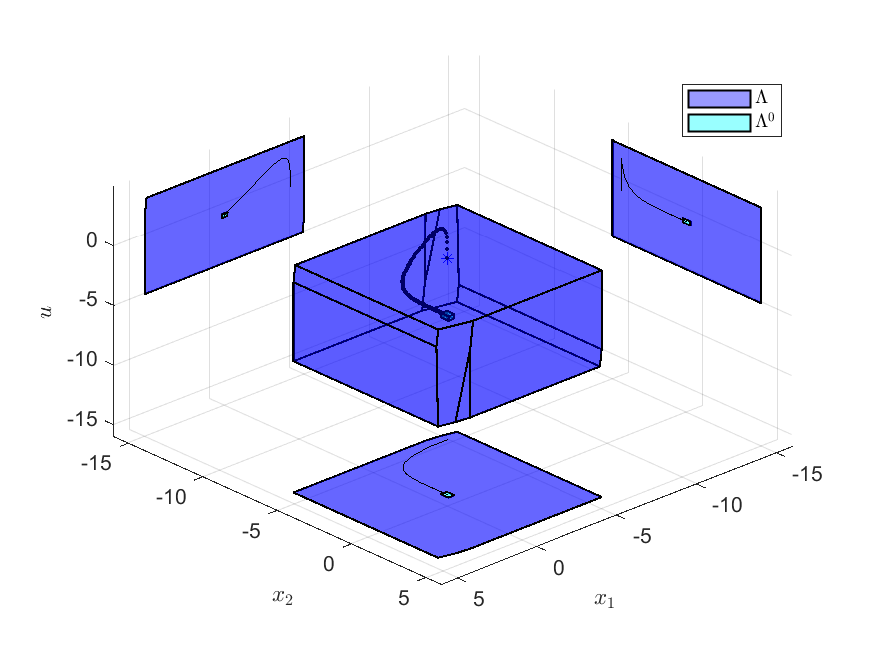}
    \caption{Example 2 - $\Lambda$ and $\Lambda^0$ sets, and system trajectory}
    \label{fig:coupledtank_set}
\end{figure}

\begin{figure}
    \centering
    \includegraphics[width=0.5\linewidth]{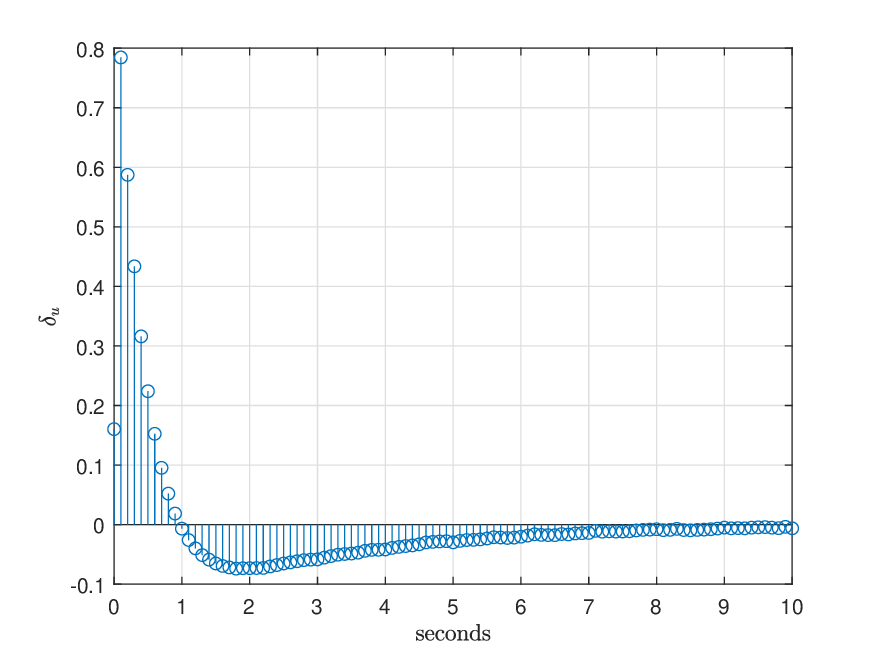}
    \caption{Example 2 - Coupled tank - Control rate}
    \label{fig:coupledtank_deltau}
\end{figure}

The resulting sets $\Lambda$ and $\Lambda^0$ are obtained from the following $\mathbb L$ and $\boldsymbol{\rho}$,

$$ \mathbb{L} =  \begin{bmatrix}
  ~~0.03744 & ~0.17477 & ~~0.00286 \\
 -0.20000 & ~~0.00000 & ~~0.00000 \\
 -0.03741 &-0.17477 &-0.00286 \\
  ~~0.0000 &-0.20000 & ~~0.00000 \\
  ~~0.00000 & ~~0.00000 & ~0.25000 \\
  ~~0.00000 & ~0.20000 & ~~0.00000 \\
  ~0.20000 & ~~0.00000 & ~~0.00000 \\
 -0.19202 & ~~0.00000 &-0.01522 \\
 -0.02041 &-0.18674 &-0.00015 \\
  ~~0.00000 & ~~0.00000 &-0.25000 \\
  ~0.19202 & ~~0.00000 & ~~0.01522 \\
  ~~0.02041 & ~0.18674 & ~~0.00015 
\end{bmatrix}, \boldsymbol{\rho} = \begin{bmatrix}
0.05028 \\ 0.03775 \\ 0.05028 \\ 0.05027 \\ 0.04236 \\ 0.05027 \\ 0.03775 \\ 0.03415 \\ 0.05055 \\ 0.04236 \\ 0.03415 \\ 0.05055 
\end{bmatrix}$$
and the resulting control gains in the form of \eqref{eq:actual-control} are given in Table \ref{tab:control_tanks}. Figure \ref{fig:coupledtank_set} illustrates the resulting sets, and their projections, where the blue \textcolor{blue}{$*$} represent the initial conditions of the system; the black bold dots represents the system state evolutions through time, and its projections represented with a black line for vision clarity. Furthermore, Figure \ref{fig:coupledtank_deltau} represents the control variation overtime, associated with the system trajectory depicted in Figure \ref{fig:coupledtank_set}, where the blue dashed line represent its bounds.

The $\Lambda$ set depicted in Figure \ref{fig:coupledtank_set} has a volume of $793.8872$, meaning it occupies $99.24 \%$ of total volume of $800$, with a projection area in $x_1$ $x_2$ of $99.5904$. Finally the inner set $\Lambda^0$ has a volume of $0.0605$, meaning less than $0.01\% $ of the total volume. All volumes were computed using the volume function of the MPT3 \cite{MPT3} toolbox.


\begin{table}[]
    \centering
    \begin{tabular}{|c|c|c|c|}
    \hline
        $i$ & $K_i$ & $\bar{K}_i$ & $\hat{K}_i$ \\ \hline \hline
        $1$ & $\begin{bmatrix}
            -1.02 &~~0.08
        \end{bmatrix}\times 10^{-5}$ & $-0.2496$ & $\begin{bmatrix}
            -0.20 & ~0.00
        \end{bmatrix}$
        \\
        $2$ & $\begin{bmatrix}           -1.02 &~~0.08 \end{bmatrix}\times 10^{-5}$ & $-0.2496$ & $\begin{bmatrix} -0.20 & ~0.00        \end{bmatrix}$ 
        \\
        $3$ & $\begin{bmatrix}
            -9.38 &-0.20
        \end{bmatrix}\times 10^{-4}$ & $-0.2496$ & $\begin{bmatrix}
            -0.20 & ~0.00
        \end{bmatrix}$ 
        \\
        $4$ & $\begin{bmatrix}
             ~3.41 &-0.018
        \end{bmatrix}\times 10^{-2}$ & $-0.2526$ & $\begin{bmatrix}
            -0.20 & ~0.00
        \end{bmatrix}$ \\ \hline
    \end{tabular}
    \caption{Example 2 - Control Gains}
    \label{tab:control_tanks}
\end{table}

\section{Conclusion}

We have developed a new output feedback design technique that effectively manages constrained discrete-time Linear Parameter-Varying (LPV) systems subject to persistent disturbances. In addition to addressing state and control constraints, this technique also accommodates control rate constraints, which frequently arise in practical control scenarios. We formulated a bilinear optimization problem based on the proposed polyhedral robust positive invariance and set inclusion conditions. This design allows for the synthesis of an incremental LPV control law that explore certain degrees of freedom not previously addressed in the existing literature. Numerical examples illustrate  the effectiveness of the proposed technique.

\section*{Appendix A}
\textbf{Proof of Lemma \ref{lem:cl-loop-Big}}:   
 We can re-write $ \mathbb{A}^{cl}(\alpha_+,\alpha) =$
 $$  \underbrace{\left[
    \begin{array}{cc}
       A(\alpha)  & B(\alpha) \\ K(\alpha)C & I+\bar{K}(\alpha) 
    \end{array}
    \right]}_{F_1(\alpha)}
    \oplus 
    \underbrace{\left[
    \begin{array}{c}
       0 \\ 
       \hat K(\alpha_+)
    \end{array}
    \right]}_{M_1(\alpha_+)}
      \underbrace{  \left[
    \begin{array}{cc}
       CA(\alpha) & CB(\alpha)
    \end{array}
    \right]}_{N_1(\alpha)}
    $$
    and $\mathbb{B}^{cl}_d(\alpha_+,\alpha)=$
    $$
    \underbrace{
    \left[
    \begin{array}{ccc}
       B^p(\alpha) & 0 & 0\\
       0 & K(\alpha)D^{\eta} & 0
    \end{array}
    \right]
    }_{F_2(\alpha)}
    \oplus 
    \underbrace{
     \left[
    \begin{array}{c}
       0 \\ 
       \hat K(\alpha_+)
    \end{array}
    \right]
    }_{M_2(\alpha_+)}
    \underbrace{
        \left[
    \begin{array}{ccc}
       CB^p(\alpha) & 0 & D^{\eta}
    \end{array}
    \right]
    }_{N_2(\alpha)}.
    $$
 
 Then, by referring  to \eqref{eq:composed_sum} with $\beta =\alpha_{+}$ and $\theta=\alpha$, we have
$$\mathbb{A}^{cl}(\alpha_+,\alpha)=
\Gamma'(\alpha_{+}) \overbrace{(\mathcal{I}^c\mathcal{F}_1 +\mathcal{M}_1^c\mathcal{N}_1)}^{\mathcal{A}^{cl}}\Gamma(\alpha),$$
 $$
\mathbb{B}^{cl}  (\alpha_+,\alpha) = \Gamma'(\alpha_+)  \overbrace{(\mathcal{I}^c\mathcal{F}_2 +\mathcal{M}_2^c\mathcal{N}_2)}^{ \mathcal{B}^{cl}}\Gamma(\alpha).$$
$\Box$

\textbf{Proof of Corollary \ref{lem:deltau-Big}}: 
The parameter-varying matrices in \eqref{eq:incrementFG}
 can be re-written
 as 
$$\mathbb A^{\delta_u}(\alpha_+,\alpha) = 
  \underbrace{\left[
    \begin{array}{cc}
 K(\alpha)C & \bar{K}(\alpha) 
    \end{array} \right]}_{F_3(\alpha_{k})}
    \oplus \underbrace{
       \hat K(\alpha_+)
}_{M_3(\alpha_+)}
      \underbrace{  \left[
    \begin{array}{cc}
       CA(\alpha) & CB(\alpha)
    \end{array}
    \right]}_{N_3(\alpha)},
    $$  
    $$\mathbb B^{\delta_u}(\alpha_+,\alpha) =
    \underbrace{
    \left[
    \begin{array}{ccc}
       0 & K(\alpha)D^{\eta} & 0
    \end{array}
    \right]
    }_{F_4(\alpha)}
    \oplus 
    \underbrace{
       \hat K(\alpha_+)
    }_{M_4(\alpha_+)}
    \underbrace{
        \left[
    \begin{array}{ccc}
       CB^p(\alpha) & 0 & D^{\eta}
    \end{array}
    \right]
    }_{N_4(\alpha)}.
    $$

Then, by referring again to \eqref{eq:composed_sum}, we obtain 
$$\mathbb{A}^{\delta_u}(\alpha_+,\alpha)
= \Gamma'(\alpha_+) \overbrace{(\mathcal{I}^c\mathcal{F}_3 +\mathcal{M}_3^c\mathcal{N}_3)}^{\mathcal{A}^{\delta_u}}\Gamma(\alpha),$$
$$
\mathbb{B}^{\delta_u}(\alpha_+,\alpha)= \Gamma'(\alpha_+) \overbrace{(\mathcal{I}^c\mathcal{F}_4+\mathcal{M}_4^c\mathcal{N}_4)}^{\mathcal{B}^{\delta_u}}\Gamma(\alpha).$$
$\Box$

\textbf{Proof of Theorem \ref{prop:RPI-conditiona}:} The proof is divided into three parts, as follows.\\ \textbf{1st - RPI with $\lambda-$contractivity of $\Lambda$}: It consists in showing that the  relations \eqref{nonlinear1}-\eqref{cond2} are equivalent to the following one-step admissibility condition that characterizes the RPI, with $\lambda$-contractivity, of $\mathbb L$ for the closed-loop system \eqref{eq:clsystem}
\begin{equation}
\label{eq:onestepRPI}
\begin{array}{c} \mathbb L \xi_+ :=
 \mathbb L \begin{bmatrix}
    \mathbb A^{cl} (\alpha_+,\alpha) & \mathbb B^{cl}(\alpha_+,\alpha)
\end{bmatrix}\begin{bmatrix}
    \xi \\ d
\end{bmatrix} \leq \lambda \mathbf 1_{l_r}, \\
\forall \xi \text{ and } d \text{ such that }  
\begin{bmatrix}
    \mathbb L & 0 \\ 0 & \mathbb D
\end{bmatrix}\begin{bmatrix}
    \xi \\ d
\end{bmatrix} \leq \begin{bmatrix}
    \mathbf 1_{l_r} \\ \mathbf 1_{l_d}
\end{bmatrix}.
\end{array}
\end{equation}
By resorting to the notation \eqref{eq:cte-premultiply-LPV}-\eqref{eq:cte-postmultiply-LPV}, the pre- and post-multiplication of each relation \eqref{nonlinear1}-\eqref{cond1} by compatible $\Gamma'(\alpha_+)$ and $\Gamma(\alpha_{k})$,  yields
\begin{subequations}
    \label{const-alpha}
    \begin{align}
H(\alpha_+,\alpha) \mathbb L & =  \mathbb L\,\mathbb A^{cl}(\alpha_+,\alpha), \label{nonlinear1-alpha}\\
V (\alpha_+,\alpha)\mathbb D &= \mathbb L\,\mathbb B^{cl}(\alpha_+,\alpha), \label{cond4-alpha} \\
 H(\alpha_+,\alpha)\,\mathbf{1}_{l_r} + V(\alpha_+,\alpha)\,\textbf{1}_{l_d} &\leq \lambda\,\textbf{1}_{l_r}, \label{cond1_alpha} \\
 H(\alpha_+,\alpha) \boldsymbol{\rho} +  V(\alpha_+,\alpha)\,\mathbf 1_{l_d} &\leq \epsilon_1 \boldsymbol{\rho}, \label{cond2-alpha}
    \end{align}
\end{subequations}
where the corresponding $H(\alpha_+,\alpha)  = \Gamma'(\alpha_+) \mathcal H \Gamma(\alpha) \in \Re^{l_r \times l_r}$ and  $V(\alpha_+,\alpha) = \Gamma'(\alpha_+) \mathcal V \Gamma(\alpha) \in \Re^{l_r \times l_d}$ are, by construction, nonnegative matrices for every $(\alpha_+,\alpha) \in \mathcal S \times \mathcal S$. Conversely, we require the relations \eqref{nonlinear1}-\eqref{cond2} hold true for the infinite dimensional relations \eqref{nonlinear1-alpha}-\eqref{cond2-alpha} be verified for all $(\alpha_+,\alpha) \in \mathcal S \times \mathcal S$.

Thus, \eqref{nonlinear1-alpha} and \eqref{cond4-alpha}, can be re-written as
$\begin{bmatrix}
     H(\alpha_+,\alpha) & V(\alpha_+,\alpha)
\end{bmatrix}
\begin{bmatrix}
    \mathbb L & 0 \\ 0 & \mathbb D
\end{bmatrix} = \mathbb L \begin{bmatrix}
    \mathbb A^{cl} (\alpha_+,\alpha) & \mathbb B^{cl}(\alpha_+,\alpha)
\end{bmatrix},$
which, together with \eqref{cond1_alpha} and by resorting to the EFL, allow us to conclude that the one-step admissibility condition \eqref{eq:onestepRPI} is verified for all $(\alpha_+,\alpha) \in \mathcal S \times \mathcal S$.

\textbf{2nd - UB of $\Lambda^0$}: As in the previous step, we can show that the relations \eqref{nonlinear1}, \eqref{cond4} and \eqref{cond2} are equivalent  to the following one-step admissibility condition that proves  the RPI of the inner set $\Lambda^0$ for the closed-loop system \eqref{eq:clsystem}, 
\begin{equation}
\label{eq:onestepRPI-L0}
\begin{array}{c} 
\mathbb L \begin{bmatrix}
    \mathbb A^{cl} (\alpha_+,\alpha) & \mathbb B^{cl}(\alpha_+,\alpha)
\end{bmatrix}\begin{bmatrix}
    \xi \\ d_k
\end{bmatrix} \leq \epsilon \boldsymbol{\rho}, \\
\forall \xi \text{ and } d_k \text{ such that }  
\begin{bmatrix}
   \mathbb L & 0 \\ 0 & \mathbb D
\end{bmatrix}\begin{bmatrix}
    \xi \\ d_k
\end{bmatrix} \leq \begin{bmatrix}
    \mathbf 1_{l_r} \\ \mathbf 1_{l_d}
\end{bmatrix}.
\end{array}
\end{equation}
Hence, any closed-loop trajectory that reaches or emanates from $\Lambda^0$ will remain ultimately bounded inside it.

\textbf{Finite-time convergence}: Finally, to show the finite-time convergence of the closed-loop trajectories starting from $\Lambda$ to $\Lambda_0$, 
consider the set $\eta \Lambda_0 = \{ \xi : \mathbb{L} \xi \leq \eta \boldsymbol{\rho} \}$, where $0 < \eta \in \Re$ is the smallest scalar such that $\Lambda_0 \in \Lambda \subseteq \eta \Lambda_0$. Notice that $\eta \Lambda_0$ is also an RPI set of the system \eqref{eq:systemplant} and shares the guaranteed contractivity coefficient $\tilde{\lambda}=\epsilon_1 < 1$ of $\Lambda_0$, with $\epsilon_1 \longrightarrow 1$. 
Thus, proceeding as in \cite{franklineEugenio2021}, for any $\xi_0 \in \Lambda \subseteq \eta \Lambda_0$ and for $k \geq \tilde{k} = \log_{\tilde{\lambda}} \eta_0 \Rightarrow \xi_k \in \Lambda_0$. 
Thus, the number $\tilde{k}$ can be seen as a worst-case upper bound for the finite number of steps to reach $\Lambda_0$ from $\Lambda$.
$\Box$


\textbf{Proof of Lemma \ref{lem:deltau-admissible}} It follows the same rationale as in the first step of the proof of Proposition \ref{prop:RPI-conditiona}. 
Thus, from pre- and post-multiplication of each relation \eqref{nonlinear31}-\eqref{condlem3} by compatible $\Gamma'(\alpha_+)$ and $\Gamma(\alpha)$, we can obtain the following equivalent infinite-dimensional relations
\begin{subequations}
\label{const31-alpha}
\begin{align}
Q(\alpha_+,\alpha)\mathbb L & =  U_\delta\,\mathbb A^{\delta_u}(\alpha_+,\alpha), \label{nonlinear31-alpha}\\
T(\alpha_+,\alpha)\mathbb D &= U_\delta\,\mathbb B^{\delta_u}(\alpha_+,\alpha), \label{cond34-alpha} \\
 Q(\alpha_+,\alpha)\,\mathbf{1}_{l_r} + T(\alpha_+,\alpha)\,\textbf{1}_{l_d} &\leq \lambda\,\textbf{1}_{l_r}, \label{cond31_alpha}
\end{align}
\end{subequations}
where $Q(\alpha_+,\alpha)  = \Gamma'(\alpha_+) \,\mathcal Q \,\Gamma(\alpha) \in \Re^{l_{u_\delta} \times l_r}$ and  $T(\alpha_+,\alpha) = \Gamma'(\alpha_+) \, \mathcal T \Gamma(\alpha)\, \Re^{ l_{u_\delta}\times l_d }$ are, by construction, nonnegative matrices for all $(\alpha_+,\alpha) \in \mathcal S \times \mathcal S$. Hence, by resorting to the EFL, the above relations \eqref{const31-alpha} are equivalent to the admissibility condition \eqref{eq:deltaufulfil_1}, as required.
 $\Box$

\bibliography{autosam.bib}           %

\begin{thebibliography}{10}
\providecommand \doibase [0]{http://dx.doi.org/}%

\bibitem{tarbouriech2011stability}
Tarbouriech S, Garcia G, {da Silva Jr} JMG, Queinnec I. {\it Stability and stabilization of linear systems with saturating actuators}.
\newblock Springer Science \& Business Media, 2011.

\bibitem{Blanchini15}
Blanchini F, Miani S. {\it Set-Theoretic Methods in Control}.
\newblock Boston: Birkhäuser, 2015.

\bibitem{mohammadpour2012control}
Mohammadpour J, Scherer CW. {\it Control of linear parameter varying systems with applications}.
\newblock Springer Science \& Business Media, 2012.

\bibitem{Tanaka02}
Tanaka K, Wang HO. {\it Fuzzy Control Systems Design and Analysis: A Linear Matrix Inequality Approach}.
\newblock New York: John Wiley \& Sons, Inc., 2002.

\bibitem{KCLS:15}
Klug M, Castelan EB, Leite VJ, Silva LF. Fuzzy dynamic output feedback control through nonlinear Takagi–Sugeno models. {\it Fuzzy Sets and Systems.} 2015\string;263\string:92-111.
\newblock \href {\doibase https://doi.org/10.1016/j.fss.2014.05.019} {doi: https://doi.org/10.1016/j.fss.2014.05.019}

\bibitem{yang2016optimal}
Yang W, Gao J, Feng G, Zhang T. An optimal approach to output-feedback robust model predictive control of LPV systems with disturbances. {\it International Journal of Robust and Nonlinear Control.} 2016\string;26(15)\string:3253--3273.

\bibitem{ping2021tube}
Ping X, Yao J, Ding B, Li Z. Tube-based output feedback robust MPC for LPV systems with scaled terminal constraint sets. {\it IEEE Transactions on Cybernetics.} 2021\string;52(8)\string:7563--7576.

\bibitem{ding2019output}
Ding B, Dong J, Hu J. Output feedback robust MPC using general polyhedral and ellipsoidal true state bounds for LPV model with bounded disturbance. {\it International Journal of Systems Science.} 2019\string;50(3)\string:625--637.

\bibitem{sereni2022stabilization}
Sereni B, Assun{\c{c}}{\~a}o E, Teixeira MCM. Stabilization and disturbance rejection with decay rate bounding in discrete-time linear parameter-varying systems via $\mathscr{H}_\infty$ gain-scheduling static output feedback control. {\it International Journal of Robust and Nonlinear Control.} 2022\string;32(14)\string:7920--7945.

\bibitem{peixoto2021improved}
Peixoto ML, Coutinho PH, Palhares RM. Improved robust gain-scheduling static output-feedback control for discrete-time LPV systems. {\it European Journal of Control.} 2021\string;58\string:11--16.

\bibitem{oliveira2021exponential}
Oliveira LA, Barbosa MV, Silva LF, Leite VJ. Exponential stabilization of LPV systems under magnitude and rate saturating actuators. {\it IEEE Control Systems Letters.} 2021\string;6\string:1418--1423.

\bibitem{da2003local}
Da~Silva JG, Tarbouriech S, Garcia G. Local stabilization of linear systems under amplitude and rate saturating actuators. {\it IEEE Transactions on Automatic Control.} 2003\string;48(5)\string:842--847.

\bibitem{kose2002rate}
K{\"o}se IE, Jabbari F. Rate and magnitude-bounded actuators: Scheduled state feedback design. {\it IFAC Proceedings Volumes.} 2002\string;35(1)\string:73--78.

\bibitem{hempel2011output}
Hempel AB, Kominek AB, Werner H. Output-feedback controlled-invariant sets for systems with linear parameter-varying state transition matrix. In: IEEE Conference on Decision and Control and European Control Conference. IEEE.  2011\string:3422--3427.

\bibitem{dorea2020robust}
D{\'o}rea CE, Castelan EB, Ernesto JG. Robust positively invariant polyhedral sets and constrained control using fuzzy ts models: a bilinear optimization design strategy. {\it IFAC-PapersOnLine.} 2020\string;53(2)\string:8013--8018.

\bibitem{ITC:23}
Isid{\'o}rio ID, D{\'o}rea CE, Castelan EB. Observer-based output feedback control using invariant polyhedral sets for fuzzy T--S models under constraints. {\it Journal of Control, Automation and Electrical Systems.} 2023\string;34(4)\string:752--765.

\bibitem{bender2012output}
Bender FA, {da Silva Jr} JMG. Output feedback controller design for systems with amplitude and rate control constraints. {\it Asian Journal of Control.} 2012\string;14(4)\string:1113--1117.

\bibitem{da2008dynamic}
{Da Silva Jr} JMG, Lim{\'o}n D, Alamo T, Camacho EF. Dynamic output feedback for discrete-time systems under amplitude and rate actuator constraints. {\it IEEE Transactions on Automatic Control.} 2008\string;53(10)\string:2367--2372.

\bibitem{franklineEugenio2021}
Bri{\~a}o SL, Castelan EB, Camponogara E, Ernesto JG. Output feedback design for discrete-time constrained systems subject to persistent disturbances via bilinear programming. {\it Journal of the Franklin Institute.} 2021\string;358(18)\string:9741--9770.

\bibitem{ECSC:21}
Ernesto JG, Castelan EB, {dos Santos} GAF, Camponogara E. Incremental output feedback design approach for discrete-time parameter-varying systems with amplitude and rate control constraints. In: 2021 IEEE International Conference on Automation/XXIV Congress of the Chilean Association of Automatic Control (ICA-ACCA).  2021\string:1-7

\bibitem{ECSC:21alt}
Ernesto JG, Castelan EB, Lucia W, {dos Santos} GAF. Alternative implementation to an incremental output-feedback design approach for constrained discrete-time parameter-varying systems. {\it IFAC-PapersOnLine.} 2022\string;55(35)\string:25--30.

\bibitem{ECL:24}
Ernesto JG, Castelan EB, Lucia W. Control-rate Constrained Output Feedback Design for LPV Systems subject to Bounded Disturbances. In: Congresso Brasileiro de Autom{\'a}tica-CBA. SBA.  2024; Rio de Janeiro, Brazil.

\bibitem{H:89}
Hennet JC. Une extension du Lemme de Farkas et son application au probleme de r{\'e}gulation lin{\'e}aire sous contraintes. {\it CR Acad. Sci. Paris.} 1989\string;308(I)\string:415--419.

\bibitem{hennet95}
Hennet JC. Discrete time constrained linear systems. {\it Control and Dynamic Systems.} 1995\string;71\string:157--214.

\bibitem{knitro06}
Byrd RH, Nocedal J, Waltz RA. {\it Knitro: An Integrated Package for Nonlinear Optimization}.
\newblock In: Large-Scale Nonlinear Optimization.Boston: Springer, 2006.

\bibitem{MPT3}
Herceg M, Kvasnica M, Jones C, Morari M. {Multi-Parametric Toolbox 3.0}. In: Proc.~of the European Control Conference.  2013; Z\"urich, Switzerland\string:502--510.
\newblock \url{http://control.ee.ethz.ch/~mpt}.

\end{thebibliography}

\end{document}